%% file: SnowmassBook-TopicalGroup.tex
\documentclass{tcibook}
\usepackage{fancyhea}
\usepackage{work}
\usepackage{bm}       
\usepackage{graphicx}
\usepackage{multirow}
\usepackage{lineno}
\usepackage{threeparttable}
\usepackage[normalem]{ulem}
\usepackage{color}

\usepackage[colorlinks = true,
            linkcolor = blue,
            urlcolor  = blue,
            citecolor = blue,
            anchorcolor = blue]{hyperref}

\input workshopsymbols.tex     

\def\authorlist#1#2{
    \vskip 0.4in
\begin{center}\begin{large} {\bf  #1 } \end{large}
    \vskip 0.2in
              #2
     \vskip 0.2in
   \end{center}
}

\begin{document}


\pagenumbering{roman}

\parindent=0pt
\parskip=8pt
\setlength{\evensidemargin}{0pt}
\setlength{\oddsidemargin}{0pt}
\setlength{\marginparsep}{0.0in}
\setlength{\marginparwidth}{0.0in}
\marginparpush=0pt


\pagenumbering{arabic}

\renewcommand{\chapname}{chap:intro_}
\renewcommand{\chapterdir}{.}
\renewcommand{\arraystretch}{1.25}
\addtolength{\arraycolsep}{-3pt}

\include{Engagement/recommendations}

\input Engagement/CommF07/Society.tex


\end{document}

%% file: workshopsymbols.tex
\def\beq{\begin{equation}}
\def\eeq#1{\label{#1}\end{equation}}
\def\eeqn{\end{equation}}

\newenvironment{Eqnarray}%
   {\arraycolsep 0.14em\begin{eqnarray}}{\end{eqnarray}}
\def\beqa{\begin{Eqnarray}}
\def\eeqa#1{\label{#1}\end{Eqnarray}}
\def\eeqan{\end{Eqnarray}}

\let\bar=\overbar

\def\lsim{\mathrel{\raise.3ex\hbox{$<$\kern-.75em\lower1ex\hbox{$\sim$}}}}
\def\gsim{\mathrel{\raise.3ex\hbox{$>$\kern-.75em\lower1ex\hbox{$\sim$}}}}

\def\del{\partial}
\def\Dslash{\not{\hbox{\kern-4pt $D$}}}
\def\dslash{\not{\hbox{\kern-2pt $\del$}}}
\def\pslash{\not{\hbox{\kern-2pt $p$}}}
\def\ETmiss{\not{\hbox{\kern-4pt $E$}}_T}

\def\Dlr{\mathrel{\raise1.5ex\hbox{$\leftrightarrow$\kern-1em\lower1.5ex\hbox{$D$}}}}

\def\MSB{{\bar{M \kern -2pt S}}}
\def\msb{{\bar{\scriptsize M \kern -1pt S}}}

\def\drb{{\bar{\scriptsize D \kern -1pt R}}}



%% file: Engagement/recommendations.tex
\newcommand{\cefgroup}{7} 
\newenvironment{recs}
    {\noindent \begin{minipage}{\textwidth} \rule{\textwidth}{2mm}}
    {\rule{\textwidth}{2mm} \end{minipage}}

\newcommand{\rec}[3]{
\nobreak \noindent \textbf{CEF0\cefgroup~Recommendation #1 -- #2}\\ 
\nobreak \rule{\textwidth}{0.4mm}
\nobreak #3\\
}

%% file: Engagement/CommF07/Society.tex
\newcommand{\C}{$^\circ$C }

\setcounter{chapter}{6} 


\chapter{Report of the Topical Group on Environmental and Societal Impacts of Particle Physics for Snowmass 2021}

\authorlist{K. Bloom, V. Boisvert, M. Headley}
   {}

\section{Introduction}
\input{Engagement/CommF07/intro}

\section{Impacts on climate}

\input{Engagement/CommF07/climate}

\section{Impacts on local communities}
\input{Engagement/CommF07/community}

\section{Impacts on non-proliferation}
\input{Engagement/CommF07/nonproliferation}

\section{Acknowledgments}
\input{Engagement/CommF07/Ack}







%% file: Engagement/CommF07/intro.tex
Particle physics, as a discipline and as a community of scientists, does not exist in a vacuum.  The choices that we make as a field and as individual scientists have an impact on on the physical environment and on the human society that we are embedded in.  The natural scale of particle-physics activities -- experimental facilities that can span miles, collaborations that can number in the thousands of scientists, and projects that can stretch over decades -- have the potential to have an out-sized impact on the physical world and on the communities that they interact with beyond particle physics.  These interactions must be intentional, respectful, and sustainable to ensure the long-term success of our field.

As part of the US Community Study on the Future of Particle Physics, {\it i.e.} the Snowmass process, a topical group on ``Environmental and Societal Impacts" was formed to study these issues, and develop recommendations to help improve our relationships with the environment and with society.  The topical group was formed relatively late in the Snowmass process, in March 2021, in response to a perceived need to address a set of issues that were not naturally covered by other topical groups within the Community Engagement Frontier.  The topical group held a kick-off town hall in July 2021~\cite{kickoff}, and held two subsequent workshops, one specifically focused on carbon emissions due to particle physics activities~\cite{ClimateWorkshop} and another focused particle physics interactions with local communities~\cite{CommunityWorkshop}.  A total of five white papers were submitted for consideration by the topical group.  Three were related to the environmental impacts of particle physics~\cite{ClimatePaper,GreenILC,AcceRD}, another provided case studies of interactions of different laboratories with their local communities~\cite{community}, and the final one discussed how particle physics experiments can be used to support nuclear non-proliferation efforts~\cite{Nuclear}.  In the following sections, we provide a summary of these papers and the recommendations they made to help ensure the long-term success of our field.

%% file: Engagement/CommF07/climate.tex
As explained in~\cite{ClimatePaper}, global climate change, and how to mitigate it, is one of the most crucial issues facing humanity today.  The Intergovernmental Panel on Climate Change (IPCC) has stated, ``It is unequivocal that human influence has warmed the atmosphere, ocean and land. Widespread and rapid changes in the atmosphere, ocean, cryosphere and biosphere have occurred''. The IPCC further states that ``Global warming of 1.5\C and 2\C will be exceeded during the 21st century unless deep reductions in CO$_2$ and other greenhouse gas emissions occur in the coming decades.''  To limit the amount of warming, we will have to achieve significant reductions in CO$_2$ emissions, to the point of net-zero emissions, along with reductions in the production of other greenhouse gases (GHGs), in line with the goals of the Paris Agreement. The 6th IPCC assessment report estimates a total budget of 300 gigatons CO2e (CO$_2$ equivalent) emissions for an 83\% chance to limit global warming to below 1.5 degrees Celsius. This amounts to about 1.1~t CO2e emissions per capita per year until 2050. This should be compared to current per capita per year emission rates of 14.2~tCO$_2$ in the United States.  Significant reductions in carbon emissions must be achieved for the future health of the planet, and measures must be taken now rather than later to begin to limit the long-term impacts.

The U.S is a crucial player in the climate agenda. It is the top producer and consumer of both oil and natural gas, it has the world’s second largest number of coal-fired power plants, and fossil fuels contributed 63\% to its overall electricity generation. On the other hand, it also has the largest nuclear and second largest renewable capacity in the world. Not only is the U.S. the second largest emitter of GHGs, but over the course of history it has cumulatively produced more than any other country. Its citizens have emissions footprints that are roughly three times the global average. The U.S also has a large impact on climate policy. For example, the success of the 2015 Paris agreement was due in significant part to the leadership demonstrated by the U.S. at that time. More recently, the current administration has submitted an updated Nationally Determined Contribution to cut greenhouse gas emissions 50-52\% below 2005 levels by 2030 and has pledged to achieve net-zero emissions by “no later than 2050”.

Current and future activities in particle physics need to be considered in this context.  The pursuit of particle physics requires substantial construction projects; the consumption of electricity in the operation of accelerators, detectors, and computing; the use of GHGs in particle detectors; and in some cases significant amounts of travel.  All of these lead to the potential for particle physicists to have a carbon impact well above that of typical citizens, and thus particle physicists should be paying attention to the impacts of the discipline on the planet and seeking to reduce them. Just as our field currently demonstrates world leadership in international cooperation towards common goals, we can also demonstrate world leadership in this critical area that impacts the future of society.

In this section we examine several contexts in which the practice of particle physics has impacts on the climate.  These include the construction of large-scale experimental facilities, the design and operation of particle detectors that make use of GHGs, the operation of computing facilities, and the common research activities of physicists, including long-distance travel.  We conclude with a set of recommendations for how we as a field can begin to reduce our impact on the climate.

\subsection{Impacts of facility construction}
\input{Engagement/CommF07/summary_facility}

\subsection{Impact of detector gases}
\input{Engagement/CommF07/summary_gases}

\subsection{Impact of computing}
\input{Engagement/CommF07/summary_computing}

\subsection{Impact of laboratories and universities}
\input{Engagement/CommF07/summary_activities}

\subsection{Recommendations}
\input{Engagement/CommF07/recs}

%% file: Engagement/CommF07/summary_facility.tex
\label{sec:facility}
A key goal of the Snowmass process is to identify promising opportunities to address the most important questions in particle physics.  We expect that these opportunities will require the construction of new, large-scale experimental facilities.  The building construction industry currently contributes 10\% of the world's total carbon emissions.  If we assume that the electric grid is successfully de-carbonized by 2040, a goal of many climate plans (for example the U.S. ``has set a goal to reach 100 percent carbon pollution-free
electricity by 2035"), then construction, rather than operations, may well dominate the climate impact of a new particle physics facility.  Here we consider the climate impact of the construction of a new accelerator facility and put it in its global context.

\subsubsection{Example calculation: FCC-ee}

One potential new energy-frontier accelerator facility foreseen by the particle physics community is the Future Circular Collider (FCC).  This accelerator would likely first collide electrons and positrons to make precision measurements and later could accelerate hadrons with a center-of-mass energy of 100 TeV, allowing for the next stage of discovery following the HL-LHC era. The FCC-ee project would operate in the era of de-carbonized electricity.  The tunnel for the accelerator would be one of the longest tunnels in the world, projected at 97.75~km circumference in the conceptual design report.  In addition, 
many bypass tunnels, access shafts, large experimental caverns, and  new surface sites are planned.  

As explained in~\cite{ClimatePaper}, a top-down estimate of the impact is obtained using rules of thumb from studies of previous road tunnel construction.  These studies attempt to obtain a complete accounting of emissions. For example, they consider the impacts of fuel and electricity used in tunnel construction, along with those of the construction materials and of the possible release of methane inherent to the excavation process.  The carbon impacts depend very much on the ``rock mass quality" of the excavation site, with a range between 5,000 and 10,000~kg CO$_2$ per meter of tunnel length.  This leads to estimates of 489 to 978~kton of CO$_2$ for the main FCC-ee tunnel. Using 500~kton of CO$_2$ as a conservative estimate and dividing by the rough number of about 6000 physicists that could be contributing to this project, this amounts to about 80~t of emissions per physicist, to be compared with the target of reaching 1.1~t of emissions per person per year introduced above. Alternatively, we estimate that 6 million trees would need to be planted to absorb this amount of CO$_2$.
 
How does this compare to other forms of civil construction?  For context, we can compare with the carbon impact of typical buildings.  A study estimates that the ``embodied"  carbon emissions during building construction is 500-600~kg of CO2e per square meter; we use 550~kg CO2e as a working value.  As a sample building we take New York City's 1 World Trade Center, a prominent recently-built skyscraper, which is 94 stories tall and 3.5~Mft$^2$.  The embodied carbon in 1 WTC is thus 197~ktons, meaning that the FCC-ee main tunnel alone has a carbon impact several times as large as one of the most significant building projects in the U.S.\ in recent years.

When the complete FCC-ee, or any similar-scale particle physics facility, is considered -- the full tunnel system, the additional buildings on the site, the materials for the accelerator and detectors -- we expect that the project will have a carbon impact similar to that of the redevelopment of a neighborhood of a major city.  This implies that the environmental impact of a future facility is going to receive the same scrutiny as that of a major urban construction project.  Our field needs to be prepared for this scrutiny, by preparing to collect and analyze data on carbon impacts, and also for taking reasonable measures for the reduction of climate impacts through the development and use of low-carbon materials, with a prioritized use of reused and recycled materials.  We can already begin investments in R\&D on how to reduce our carbon impact to prepare for future environmental reviews.  

\subsubsection{The Green ILC}

As described in~\cite{GreenILC}, another proposed future collider is the International Linear Collider, which would collide electrons and positrons at various center-of-mass energies ranging from 250~GeV in an initial phase and eventually reaching 1~TeV. The total power consumption of such a collider has been estimated to range from 111~MW to 300~MW. These values are comparable to the power consumption of the LHC (180~MW) but are still large in absolute terms. The Advanced Accelerator Association in Japan, consisting of members from both industry and academia has organized a Green-ILC Working Group. Its activities includes studies on the efficient design of ILC components, accelerator sub-systems, the overall system design, and even an ILC city hosting the laboratory campus. High-efficiency components of the ILC include the recent development of a new klystron technology, which combined with modern computer tools will allow the boosting of the efficiency of the L-band klystron from around 65\% in existing ILC commercial tubes to almost 85\% in the new design. The fabrication of prototype klystrons to realize this new technology is under study now.

The Green-ILC WG also discussed the design of an ILC city that includes the ILC Laboratory campus. If the ILC is realized in Japan, it is likely that the ILC Laboratory and a surrounding new ILC city will be built near the ILC machine. In that case, the city will be newly constructed, and so advanced concepts for an efficient and sustainable city might be introduced, like for example the usage of smart power grids and a biomass power network. The biomass power network for the ILC city would include methane fermentation, biomass diesel fuel production, and scrap wood recycling factories. Biomass would be collected through the network, and various kinds of energy would be produced by biomass and distributed to residents, offices, buildings, facilities, and factories. The electric power produced by the biomass power network would be provided to the smart grid network. The ILC machine would also contribute to the biomass power network by the use of waste heat from the ILC tunnel.

Currently, the ILC is expected to emit 320 kilotons of CO$_2$ per year, compared to 871 kilotons of CO$_2$ emitted in 2018 by Ichinoseki City, the closest city to the ILC Kitakami site. Forests in this local area can absorb about 300 kilotons/year. It then is feasible for the ILC Laboratory, working with local authorities, to shape its planning to offset these losses. In particular: (1) The ILC community should develop energy-saving technologies and not only apply them to the ILC, but also give them back to society. (2) The ILC community should cooperate with the community of the area to increase the percentage of renewable energy in the area. (3) The ILC Laboratory should integrate into its construction plan a program of sound management of the local forestry industry to increase the absorption of CO$_2$.

\subsubsection{Accelerator R\&D}
As mentioned in~\cite{AcceRD}, there is a definite focus on the energy efficiency and power reduction of future large accelerators. Current research includes activities to improve the energy efficiency of accelerator components such as low loss superconducting resonators, efficient radio frequency sources, the usage of permanent magnets and highly efficient cryogenic systems. In addition, various accelerator concepts can also significantly reduce power and energy consumption. This includes ongoing research into energy recovery linacs, studies into a possible future collider which scales favourably in terms of achievable luminosity per grid power, and also ideas related to mitigating the impact of high energy colliders on the grid by actively managing their power consumption using local storage or dynamic operation. 
 
 Any facility construction project will have a measurable impact on climate change, and as responsible citizens we must consider what we can do to reduce the potential harms to future generations by our actions.

%% file: Engagement/CommF07/summary_gases.tex
As reported in~\cite{ClimatePaper}, according to CERN's environmental reports, the dominant source of CO2e emissions at the laboratory are from greenhouse gases used for detectors and for cooling. Whether the LHC experiments are in operation or in a shut down, gas emissions dominate over the emissions stemming from CERN's electricity usage. In general SF$_6$, 
hydrofluorocarbon (HFC) and perfluorocarbons (PFC) gases are used in particle detection. HFCs and PFCs are also used for detector cooling, HFCs are used in air conditioning systems, and SF$_6$ is also used for electrical insulation in power supply systems. All these gases are subject to the UN Kyoto protocol and their usage shall fade out according to the Kigali amendment of the Montreal protocol. Due to their very high GWP, F-gases are under regulation in the EU and mandatory reporting in the U.S.. Consequently, their continued procurement and price for the whole duration of the LHC program is under threat. CERN has put together various strategies to mitigate the emissions from those gases. For example, gas re-circulation is used for all gas systems across the LHC experiments. In addition, gas recuperation is also in use in some areas. During LS2, an extensive campaign of fixing leaks has occurred. 

In the longer term, both for current and future detectors, finding alternative gases with lower GWP would be very beneficial, and studies are currently ongoing along those lines. Although new liquids and gases have been developed for industry as refrigerants and high voltage insulating media, those are not necessarily appropriate for detector operation, especially taking into account the constraints of having to operate the current detectors. For example, there cannot be changes to the high voltage system or to the front-end electronics. Finding replacement gases needs to take into account several factors: their safety (non-flammable and low toxicity) and their environmental impact (low GWP) while maintaining their detector performance (including preventing the ageing of the detectors, ensuring good quenching and being radiation-hard). 

Looking to the long term future, these results highlight the crucial need to design future detectors (including cooling systems) with gas GWP in mind. Over the next few years, it will be imperative to perform R\&D aimed at reducing the GHG emissions of future detectors and cooling systems as much as possible.

%% file: Engagement/CommF07/summary_computing.tex
As discussed in~\cite{ClimatePaper}, large-scale computing is an important element of particle physics research, across both experiment and theory.  Computing is also a significant and growing source of greenhouse gas emissions through the electricity used in computation.   While data centers and computing currently contribute approximately 2-4\% of greenhouse gas emissions, this fraction is only predicted to grow in the future.

The primary tool for reducing particle physics contributions to greenhouse gas emissions through the use of computing is the choice of location for data centers.  The carbon intensity of power generation varies greatly by country and even by region within country.  When making choices about computing deployments, it is possible to go beyond considerations like computational performance per unit cost, and also consider computational performance per amount of carbon emitted through the power consumed.

On the demand side, we can be more intentional about the scheduling of computing tasks by choosing time when the computation would be less carbon intense.  This would require data centers to expose information about carbon emissions to the scheduling mechanisms.  We can also invest greater effort in optimizing computer codes, especially those used commonly in the particle physics community,  to be less computation-intense, and therefore less emissions-intense.

In the much longer term, we expect that electricity generation will be increasingly de-carbonized, as we approach the goal of a carbon-free electric grid by 2040.  This should reduce the impact of computation on climate change, but at the same time we can expect growing demand for electricity as {\it e.g.} electric cars become more prevalent.

%% file: Engagement/CommF07/summary_activities.tex
As mentioned in~\cite{ClimatePaper}, particle physics laboratories and universities are expected to have emissions associated with their research activities along with those from other work-related activities.  GHG emissions fall under three scopes widely used in the reporting of emissions: Scope 1 refers to direct emissions from the organization, Scope 2 includes indirect emissions, most notably from electricity generation, heating, etc., and Scope 3 includes all other indirect emissions, upstream and downstream of the organization, including e.g. business travel, personnel commutes, catering, etc.  For a laboratory such as CERN, Scope 1 emissions are dominated by the release of gases from detectors as discussed above.  Scope 2 emissions at both CERN and Fermilab are dominated by those associated with the accelerator complex, when the accelerators are running.  CERN gets its electric power from France, which is heavily nuclearized and 88\% carbon free.  Fermilab, by contrast, uses electrical power from sources that are only about 37\% carbon free, and purchases renewable energy certificates to offset its Scope 2 emissions.  Starting in 2024, when the PIP-II accelerator complex comes online, to be followed by LBNF, Fermilab's energy consumption is expected to increase by 30\% over historic peak levels, and we can expect similarly increased carbon emissions unless the means of power generation changes.

Universities and other research institutes produce greenhouse gas emissions over a wide variety of activities.  Many universities in U.S. and abroad now track their emissions across all scopes, although there can be many uncertainties in the tabulation of Scope 3 emissions.  While reported numbers vary widely, many universities have indicated that their per-capita emissions are well above the 1~t CO2e per year that is necessary to stay within the carbon budget to avoid excessive global warming.

There are many issues around travel, as carbon emissions from aircraft are a rapidly growing portion of all greenhouse gas emissions.  Many experimenters make regular commutes to experiment sites, either for operations tasks or for meetings.  These could potentially be minimized through a greater use of remote control centers, either at individual institutes or regional centers, or improved videoconferencing technology that can give an equivalent experience for all meeting participants regardless of location.  Conference travel can also lead to significant emissions, especially for conferences held at remote locations; previous studies have found that 1~t CO2e emissions per conference participant is typical.  This can be mitigated in part by careful choice of conference location, or by hosting international conferences across multiple regional hubs that would minimize the total amount of travel.  However, there are also many benefits to in-person meetings for developing relationships and starting projects.  The recent global pandemic has taught us much about what can be done virtually, and what cannot be.  In general, it is best to carefully evaluate for what purposes and by whom air travel is valuable, and when travel instead can be substituted by a virtual interaction.

%% file: Engagement/CommF07/recs.tex
\label{sec:recs}
We offer the following recommendations on how we as a field can reduce our impact on the climate and moderate the ongoing trends of global climate change.

\rec{1}{New experiments and facility construction projects should report on their planned emissions and energy usage as part of their environmental assessment, which will be part of their evaluation criteria.}{These reports should be inclusive of all aspects of activities, including construction, detector operations, computing, and researcher activities.}

\rec{2}{U.S. laboratories should be involved in a review across all international laboratories to ascertain whether emissions are reported clearly and in a standardized way.}{This will also allow other U.S. particle physics research centers (including universities) to use those standards for calculating their emissions across all scopes.}

\rec{3}{Using the reported information as a guide, all participants in particle physics -- laboratories, experiments, universities, and individual researchers -- should take steps to mitigate their impact on climate change by setting concrete reduction goals and defining pathways to reaching them by means of an open and transparent process involving all relevant members of the community.}{}

\rec{4}{U.S. laboratories should invest in the development and affordable deployment of next-generation digital meeting spaces in order to minimize the travel emissions of their users.}{Moreover the particle physics community should actively promote hybrid or virtual research meetings and travel should be more fairly distributed between junior and senior members of the community. For in-person meetings, the meeting location should be chosen carefully such as to minimize the number of long-distance flights and avoid layovers.}

\rec{5}{Long-term projects should consider the evolving social and economic context, such as the expectation of de-carbonized electricity production by 2040, and the possibility of carbon pricing that will have an impact on total project costs.}{}
 
\rec{6}{All U.S. particle physics researchers should  actively engage in learning about the climate emergency and about the climate impact of particle-physics research.}{}

\rec{7}{The U.S. particle physics community should promote and publicize their actions surrounding the climate emergency to the general public and other scientific communities.}{}
    
\rec{8}{The U.S. particle physics community and funding agencies should engage with the broader international community to collectively reduce emissions.}{}

%% file: Engagement/CommF07/community.tex
\label{sec:community}

As large employers and leading entities within their communities, particle physics laboratories can benefit from community engagement focused on local impacts. Community engagement plays an essential role in local decision-making, building relationships, and important discussions about the implementation of key projects. Large particle physics projects funded by the U.S. Government require an evaluation and mitigation of each project’s potential impacts on the local communities. Beyond satisfying governmental requirements, lasting and positive change can result when laboratories work alongside their respective communities in a meaningful way, which broadens the positive societal impacts of particle physics research.

Reference~\cite{community} decribes local community engagement efforts made by three laboratories: Lawrence Berkeley National Laboratory (Berkeley Lab) located in Berkeley, California; Fermi National Accelerator Laboratory (Fermilab) located in Batavia, Illinois; and the Sanford Underground Research Facility (SURF). Each case study presents a community engagement undertaking focused on local impacts distinct to each laboratory. Further, each study highlights the unique circumstances from each of the laboratory’s regions.  Berkeley provides an urban perspective to community engagement through its Community Relations and K-12 STEM Education and Outreach programs and through partnering with a local non-profit, Rising Sun, which focuses on workforce development programs for youth. Fermilab presents an example of its suburban approach to community engagement through the work of its Community Advisory Board, while SURF highlights a rural approach to engaging with indigenous groups through education and cultural awareness through the creation of an ethnobotanical garden.

Although each of the case studies presents a different perspective on community engagement as it interfaces with social impact, several common themes emerge across all three. In all of the studies, employing consistent outreach techniques, promoting diversity, establishing lasting relationships, and creating environments for open and honest communication led to the best outcomes.

In addition to the above examples, the astronomy and cosmic communities have built research facilities on tribal lands and on sacred lands of indigenous communities. Their experiences have also demonstrated that improving relationships with indigenous communities is essential.

We offer the following recommendations to laboratories or other entities looking to begin or expand their community engagement efforts: 

\rec{9}{Laboratories should engage with their local communities in order to create awareness about their work and build lasting, positive relationships.}{Community engagement plays an essential role in local decision-making, building relationships, and important discussions about the implementation of key projects. Large particle physics projects funded by the U.S. Government require an evaluation and mitigation of each project’s potential impacts on the local communities. In addition to satisfying governmental requirements, working alongside their local communities can foster lasting change that broadens the positive societal impacts of particle physics research.}

\rec{10}{Laboratories should have consistent outreach and engagement efforts that provide regular opportunities for feedback to help establish trust.}{Through its Community Advisory Board, Fermilab offered regularly scheduled meetings to gain feedback from local communities. In addition, SURF ensured its communication with stakeholders at Isna Wica Owayawa was consistent and persistent in order to overcome scheduling and other barriers.}

\rec{11}{Laboratories should promote diversity of membership and collaborative efforts in their outreach initiatives to bring a variety of perspectives to the table and create a better end project.}{SURF’s work with tribal elders and other leaders in its local community helps ensure perspectives of indigenous populations in the region are represented and reflected in the work of the Sacred Circle Garden. Meanwhile, Fermilab regularly refreshes and expands its CAB membership to ensure it remains representative of the diversity of its suburban area.}

\rec{12}{Laboratories should avoid transactional relationships when developing relationships with stakeholders, and instead focus on approaches that provide value to each entity.}{Laboratories will be best served by making an extended commitment to working with collaborators over an extended period of time, rather than one-time interactions. Opportunities to receive feedback and consider changes can have lasting impacts on the collaborative efforts. SURF has continued to see improvement in program outcomes with Isna Wica Owayawa using this approach. Berkeley Lab has seen success by utilizing small investments in staff time, small-scale donations, and other resources as a launch pad for lasting collaborations with organizations with shared goals and values.}

\rec{13}{Laboratories should utilize methods that promote honest, two-way communication when engaging in collaborative efforts with stakeholders.}{All three case studies exemplify the benefits of open communication. The CAB at Fermilab creates a space where local community members and the lab are able to air concerns and discuss solutions. Berkeley Lab ensures that its community engagement interactions provide a space for members of the community and partners to voice their opinions, while Berkeley Lab listens and reflects on the opinions shared. Finally, SURF seeks indigenous perspectives although in some instances, the resulting dialogue can result in uncomfortable conversations. However, by promoting difficult conversations in a safe environment, SURF was able to promote a design for its ethnobotanical garden that was approved by all involved.}

%% file: Engagement/CommF07/nonproliferation.tex
As explained in~\cite{Nuclear}, in nonproliferation contexts, it is often difficult to reconcile the conflicting goals of non-intrusiveness and robust verification of the absence of illicit nuclear programs. One important area of interest is to confirm either the absence, or conversely the licit operation, of nuclear reactors -- the sources of all the world’s plutonium. Antineutrinos, whose penetrating signature of nuclear
origin can provide insight into the operations and characteristics of nuclear reactors, offer a promising path towards remote and non-intrusive but still robust and persistent verification of reactor operations. Since approximately 2010, U.S. nonproliferation researchers, supported by the National Nuclear Security Administration (NNSA), have been studying a range of possible applications of relatively large (100 ton) to very large (hundreds of kiloton) water and scintillator neutrino detectors. In parallel, the high energy physics community is pursuing similar technical goals that advance fundamental physics, in the areas of MeV scale solar, atmospheric and geological neutrinos, supernova detection, neutrinoless double beta decay, and the GeV-scale high energy neutrino oscillation and CP violation physics that are the main goals of the U.S.’s flagship Deep Underground Neutrino Experiment (DUNE). 

The following recommendations are presented:

\rec{14}{The High Energy Physics community should  continue to engage in a natural synergy in research activities into next-generation large scale water and scintillator neutrino detectors, now being studied for remote reactor monitoring, discovery and exclusion applications in cooperative nonproliferation contexts.}{}

\rec{15}{Examples of ongoing synergistic work at U.S. national laboratories and universities that support nonproliferation efforts should continue and be expanded upon.}{These include prototype gadolinium-doped water and water-based and opaque scintillator test-beds and demonstrators, extensive testing and industry partnerships related to large area fast position-sensitive photomultiplier tubes, and the development of concepts for a possible underground kiloton-scale water-based detector for reactor monitoring and technology demonstrations.}

\rec{16}{Opportunities for engagement between the particle physics and nonproliferation communities should be encouraged.}{Examples include the bi-annual Applied Antineutrino Physics conferences, collaboration with U.S. National Laboratories engaging in this research, and occasional NNSA funding opportunities supporting a blend of nonproliferation and basic science R\&D, directed at the U.S. academic community.}





%% file: Engagement/CommF07/Ack.tex
We thank the authors of the submitted white papers for all of their efforts to highlight the impacts of particle physics on the environment and on society.

%% file: SnowmassBook-TopicalGroup.bbl
\begin{thebibliography}{99}


\bibitem{kickoff} ``Kick Off Town Hall", online meeting held on 22nd July 2021, \url{https://indico.fnal.gov/event/49647/}

\bibitem{ClimateWorkshop} ``Workshop on Carbon Emissions at Future Facilities", online workshop held on 9th November 2021, \url{https://indico.fnal.gov/event/51385/}

\bibitem{CommunityWorkshop} ``Workshop on Local Community Impacts", online workshop held on 15th November 2021, \url{https://indico.fnal.gov/event/51648/}

\bibitem{ClimatePaper} Bloom, K., Boisvert, V.,  Britzger, D., Buuck, M., Eichhorn, E., Headley, M., Lohwasser, K. and Merkel, P. ``Climate impacts of particle physics",  \url{https://arxiv.org/abs/2203.12389}

\bibitem{GreenILC} Aryshev, A. et al, ``The International Linear Collider: Report to Snowmass 2021", \url{https://arxiv.org/abs/2203.07622}

\bibitem{AcceRD} Roser, T. on behalf of the ICFA Panel for Sustainable Accelerators and Colliders, ``Sustainability Considerations for Accelerator and Collider Facilities", \url{https://arxiv.org/abs/2203.07423}

\bibitem{community} Zens, R. and Headley, M. and Wolf, D. and Markovitz, A. and Dukes, F. and Tang, J. and Bloom, K. and Boisvert, V. ``Societal impacts of particle physics projects", \url{https://arxiv.org/abs/2203.07995}

\bibitem{Nuclear} Akindele, T. et al, ``A Call to Arms Control: Synergies between Nonproliferation Applications of Neutrino Detectors and Large-Scale Fundamental Neutrino Physics Experiments", \url{https://arxiv.org/abs/2203.00042} 



\end{thebibliography}
